\documentclass[fleqn]{2023SCGE}
\usepackage{graphicx}

\usepackage{colortbl}
\usepackage{xcolor}
\usepackage{float}
\usepackage{subfigure}
\usepackage{adjustbox}
\usepackage{multirow}
\usepackage{booktabs}
\usepackage{amsmath}
\usepackage{amssymb}
\usepackage{enumitem}
\usepackage{supertabular}

\usepackage{times}
\usepackage{tablefootnote}
\usepackage{overpic}
\usepackage{threeparttable}
\usepackage{hyperref}

\begin{document}

\ensubject{subject}

\ArticleType{Article}
\SpecialTopic{SPECIAL TOPIC: }
\Year{2025}
\Month{xx}
\Vol{xx}
\No{xx}
\DOI{??}
\ArtNo{000000}
\ReceiveDate{xx}
\AcceptDate{xx}

\title{Does the radio-active phase of XTE~J1810$-$197 recur following the same evolutionary pattern?}{Does the radio-active phase of XTE~J1810$-$197 recur following the same evolutionary pattern?}

\author[1]{Zhipeng Huang}{}
\author[4,2,3]{Zhen Yan}{{yanzhen@shao.ac.cn}}
\author[2,3,4]{Zhiqiang Shen}{zshen@shao.ac.cn}
\author[5]{Hao Tong}{}
\author[6,7]{Mingyu Ge}{}
\author[8,9]{Zhifu Gao}{}
\author[2,4]{\\Yajun Wu}{}
\author[2,4]{Rongbing Zhao}{}
\author[2]{Jie Liu}{}
\author[2,3]{Rui Wang}{}
\author[2,3]{Xiaowei Wang}{}
\author[2,10]{Fan Yang}{}
\author[2,3]{\\Chuyuan Zhang}{}
\author[2,3]{Zhenlong Liao}{}
\author[2,11]{Yangyang Lin}{}

\AuthorMark{Z. Huang}

\AuthorCitation{Z. Huang, Z. Yan, Z Shen, et al}

\address[1]{School of Physics and Mechanical \& Electrical Engineering, Institute of Astronomy and High Energy Physics, Hubei University of Education, Wuhan 430205, China}
\address[2]{Shanghai Astronomical Observatory, Chinese Academy of Sciences, Shanghai 200030, China}
\address[3]{School of Astronomy and Space Sciences, University of Chinese Academy of Sciences, Beijing 100049, China}
\address[4]{State Key Laboratory of Radio Astronomy and Technology, Shanghai Astronomical Observatory, Chinese Academy of Sciences, Shanghai 200030, China}
\address[5]{Department of Astronomy, School of Physics and Materials Science, Guangzhou University, Guangzhou 510006, China}
\address[6]{State Key Laboratory of Particle Astrophysics, Institute of High Energy Physics, Chinese Academy of Sciences, Beijing 100049, China}
\address[7]{University of Chinese Academy of Sciences, Chinese Academy of Sciences, Beijing 100049, China}
\address[8]{Xinjiang Astronomical Observatory, Chinese Academy of Sciences, Urumqi 830011, China}
\address[9]{State Key Laboratory of Radio Astronomy and Technology, Xinjiang Astronomical Observatory, Chinese Academy of Sciences, Urumqi 830011, China}
\address[10]{School of Physical Science and Technology, ShanghaiTech University, Shanghai 201210, China}
\address[11]{College of Science, Shanghai University, Shanghai 200444, China}

\abstract{Magnetars are the most strongly magnetized compact objects known in the Universe and are regarded as one of the primary engines powering a variety of enigmatic, high-energy transients. However, our understanding of magnetars remains highly limited, constrained by observational sample size and radiative variability.  XTE~J1810$-$197, which re-entered a radio-active phase in 2018, is one of only six known radio-pulsating magnetars. Leveraging the distinctive capability for simultaneous dual-frequency observations, we utilized the Shanghai Tianma Radio Telescope (TMRT) to monitor this magnetar continuously at both 2.25 and 8.60~GHz, capturing its entire evolution from radio activation to quenching. This enabled precise characterization of the evolution in its integrated profile, spin frequency, flux density, and spectral index ($\alpha$, defined by $S \propto f^{\alpha}$). The first time derivative of its spin frequency $\dot{\nu}$ passed through four distinct phases---rapid decrease, violent oscillation, steady decline, and stable recovery---before returning to its pre-outburst value concomitant with the cessation of radio emission. Remarkably, both the amplitudes and the characteristic time-scales of these $\dot{\nu}$ variations match those observed during the previous outburst that began in 2003, providing the first demonstration that post-outburst rotational evolution and radiative behavior in a magnetar are repeatable. A twisted-magnetosphere model can qualitatively account for this repeatability as well as for the progressive narrowing and abrupt disappearance of the radio pulse radiation, thereby receiving strong observational support.}



\maketitle


\begin{multicols}{2}
\section{Introduction}\label{sec:intro}
Magnetars are a subclass of isolated neutron stars characterized by long spin periods ($P\sim1-12$~s) and ultra-strong magnetic fields ($B\sim10^{13}-10^{15}$~G). They are primarily identified through their transient high-energy outbursts, observed as soft gamma-ray repeaters (SGRs) and anomalous X-ray pulsars (AXPs). These energetic events, which are often too powerful to be sustained by rotational energy loss, are widely interpreted as being powered by the decay of their magnetic fields \cite{dt1992}. Magnetars are rare, with only about 32 confirmed sources and candidates, and even fewer are radio-loud---only six have been observed to emit pulsed radio emission \cite{ok2014,cbi2021,enh2021}\footnote{\url{http://www.physics.mcgill.ca/~pulsar/magnetar/main.html}}. Unlike the relatively stable emission of ordinary pulsars, radio magnetars show striking variability: their pulse profiles change dramatically, rotational behavior is often irregular, flux densities vary significantly, and their spectra are typically flat or inverted \cite{crp2007,akn2013,tek2015,tde2017,crd2022}.

Unlike the relatively stable radio emission of ordinary pulsars, magnetar radio signals are transient, typically emerging after intense X-ray outbursts \cite{crh2016,scs2017,hne2024}. Constrained by this transient nature and observational limitations, earlier campaigns often failed to determine the precise onset of radio activity, thereby limiting our understanding of magnetar evolution over a complete radio-active phase. For instance, the radio emission of XTE~J1810$-$197, first detected in 2004, followed the 2003 X-ray outburst \cite{ims2004,hgb2005}, but the exact onset time of the radio activity immediately after the outburst remains undetermined due to insufficient coverage \cite{lld2019}. Similarly, the radio discovery of 1E~1547.0$-$5408 in 2007 was likely triggered by an unrecorded high-energy event \cite{crh2007,hgr2008}, a scenario also inferred for PSR~J1622$-$4950 \cite{lbb2010,ags2012}. With the advent of observational facilities such as the Swift Burst Alert Telescope (BAT), early monitoring of magnetar radio-active phases has become feasible \cite{ier2016,gha2019,egk2020}, thereby providing an opportunity to investigate magnetar evolution over the entirety of its radio-active phase.
 
XTE~J1810$-$197 was discovered through an X-ray outburst in 2003 and identified as a transient magnetar with a spin period of 5.54~s \cite{ims2004}. Based on its period derivative ($\dot{P} \sim 10^{-11}~\mathrm{s}~\mathrm{s}^{-1}$) at the time of its discovery, the source is characterized by an extremely strong magnetic field of $B \sim 3\times 10^{14}$~G and a small characteristic age of $\tau_c \lesssim 7600$~yr \cite{ims2004}. One year later, radio pulsations from this source were detected by the Very Large Array (VLA), confirming it as the first radio-loud magnetar \cite{hgb2005}. This period of activity, hereafter referred to as the 2003 outburst, spans from the 2003 X-ray outburst to the cessation of radio emission in late 2008. During the 2003 outburst, XTE~J1810$-$197 exhibited several interesting behaviors. In contrast to the stable rotation properties of ordinary pulsars, the spin frequency derivative ($\dot\nu$) of XTE~J1810$-$197 experienced a phase of decline followed by a subsequent recovery. The integrated pulse profile showed dramatic variability in both the active emission regions and their relative intensities \cite{ccr2007,ksj2007}. Uniquely among known radio magnetars, XTE~J1810$-$197 was the only magnetar where a radio interpulse (IP) has been observed. Kramer et al. (2007) first confirmed the detection of the IP in 2006 based on polarimetric observations, which they subsequently used to fit the Rotating Vector Model (RVM) \cite{ksj2007}. In most cases, XTE~J1810$-$197 showed relatively flat spectra with typical spectral indices ($\alpha$, defined by $S \propto f^{\alpha}$) greater than $-$0.8. Toward the final stage of the outburst, however, its properties evolved noticeably: profile variations became less pronounced, $\dot{\nu}$ stabilized, and the spectrum steepened significantly (approaching $\alpha \approx -3$) \cite{crh2016,lld2019}. These changes have been interpreted as a possible transition toward radio characteristics more closely resembling those of ordinary pulsars \cite{crh2016}.

After a decade of quiescence, XTE~J1810$-$197 reactivated in December 2018, hereafter the 2018 outburst, with renewed X-ray activity accompanied almost simultaneously by the reemergence of pulsed radio emission \cite{gha2019,lld2019}. XTE~J1810$-$197 exhibited irregular rotation once again, with variations in $\dot\nu$ comparable in amplitude to those seen during the 2003 outburst \cite{dlb2019,crd2022}. The integrated pulse profile was similarly unstable, showing complex structures and evolving rapidly over time with significant changes in both the number and relative strength of components \cite{crd2022,msj2022}. However, unlike the 2003 outburst, during which a bright and frequent IP was observed, no distinct IP component was detected in 2018; only a weak IP was reported at 8.60~GHz with a signal-to-noise ratio (S/N) of 3.76 \cite{hys2023}. The radio spectrum generally remained flat or slightly inverted \cite{eta2021,hys2023}, consistent with the typical properties of radio magnetars. Since the onset of the 2018 outburst, we have been conducting regular dual-frequency (2.25/8.60~GHz) observations of XTE~J1810$-$197 using the Tian Ma Radio Telescope (TMRT), which is a fully steerable 65-meter Cassegrain antenna located in Shanghai. As part of this long-term monitoring campaign, our previous study presented results from 194 epochs of simultaneous dual-frequency observations spanning 926 days (MJD~58501--59427; 2019 January 18--2021 August 1), providing a systematic description of the early-phase evolution in its integrated profiles, rotation, flux densities, and spectral indices \cite{hys2023}.

Following our earlier observations, we carried out an additional 102 epochs of simultaneous dual-frequency monitoring over a span of nearly 1000 days (MJD~59636--60612; 2022 February 26--2024 October 29), during which XTE~J1810$-$197 eventually became undetectable in the radio band. In combination with our previous dataset, this enables us to characterize the full-scale evolution of XTE~J1810$-$197 across its entire radio-active phase. This also provides an opportunity to explore potential connections between the 2003 and 2018 outbursts. The structure of this paper is as follows: in sect.~\ref{sec:obs}, we describe the observations and data reduction. Sect.~\ref{sec:result} presents the results of our observations after MJD~59427 and combines them with earlier data to illustrate the full evolution of various radio parameters throughout the 2018 outburst. In sect.~\ref{sec:discussion}, we perform a comparative analysis of the 2003 and 2018 outbursts, highlighting similarities and differences, and discuss their implications for the magnetar radio emission mechanism. Finally, sect.~\ref{sec:conc} summarizes our main findings.

\section{Observation and Data Reduction} \label{sec:obs}
As early as 2018, pioneering simultaneous pulsar observations at 2.25/8.60~GHz with the TMRT had already achieved success \cite{ysm18}. This distinctive approach of simultaneous dual-frequency observation has subsequently gained broader applications, yielding a series of significant achievements in areas such as constraining pulsar radiation and understanding interstellar medium properties \cite{wys24, wys25}. Following our previous research work \cite{hys2023}, we had been monitoring XTE~J1810$-$197 simultaneously at 2.25/8.60~GHz using the TMRT from MJD~58501 to 60612. In total, 296 dual-frequency observations were performed. The effective frequency coverage is 2.20--2.30~GHz (S-band) and 8.20--9.00~GHz (X-band), respectively.
 
The dual-frequency signals were  sampled and digitized with the Digital Backend System (DIBAS), which consists of three parallel ROACH-2/ADC processing units \cite{ysm2018}. To mitigate the effect of interstellar dispersion and to improve frequency-domain radio frequency interference (RFI) mitigation, the total bandwidth was divided into sub-channels of 1~MHz (S-band) and 2~MHz (X-band). Two observing modes were used: the online-folding mode and the search mode, both implemented within the DIBAS system by inheriting functionalities from the Green Bank Ultimate Pulsar Processing Instrument (GUPPI) \cite{drd08}. In the online-folding mode, data were incoherently de-dispersed and folded in real time using predictive timing files generated by the \texttt{TEMPO} software \footnote{\url{https://tempo.sourceforge.net/}}. Each pulse period was divided into 1024 phase bins and sub-integrated every 30~s. A dispersion measure (DM) of 178~$\mathrm{cm}^{-3}$pc was applied during folding \cite{crh2006}. In search mode, data were sampled at 131.96~$\mu$s and later processed manually using \texttt{DSPSR} \cite{vb2011}, employing the same phase resolution, sub-integration time, and DM to ensure consistency with the folding-mode data. 

All datasets were manually inspected and cleaned of RFI using the \texttt{pazi} tool from the \texttt{PSRCHIVE} software package \cite{hsm2004,sdo2012} \footnote{\url{https://psrchive.sourceforge.net/}}. As no real-time flux calibration system was available during our observations, we estimated the flux densities of XTE~J1810$-$197 using the radiometer equation, based on the observed S/N and the system equivalent flux density (SEFD) of the receiver system \cite{lrf11, zwy2017}. The SEFD of the TMRT is 46~Jy and 48~Jy at 2.25/8.60~GHz, respectively \cite{wzy2016,ysm2018}. The mean flux density was then calculated by dividing the integrated flux over the on-pulse bins by the total number of phase bins. The corresponding uncertainty was estimated from the root-mean-square (RMS) of the off-pulse region, normalized by the square root of the number of phase bins \cite{zwy2017}.

\section{Results} \label{sec:result}
\subsection{Long-term Evolution of Integrated Profiles}

\begin{figure*}[ht!]
\centering
\includegraphics[width=1\textwidth]{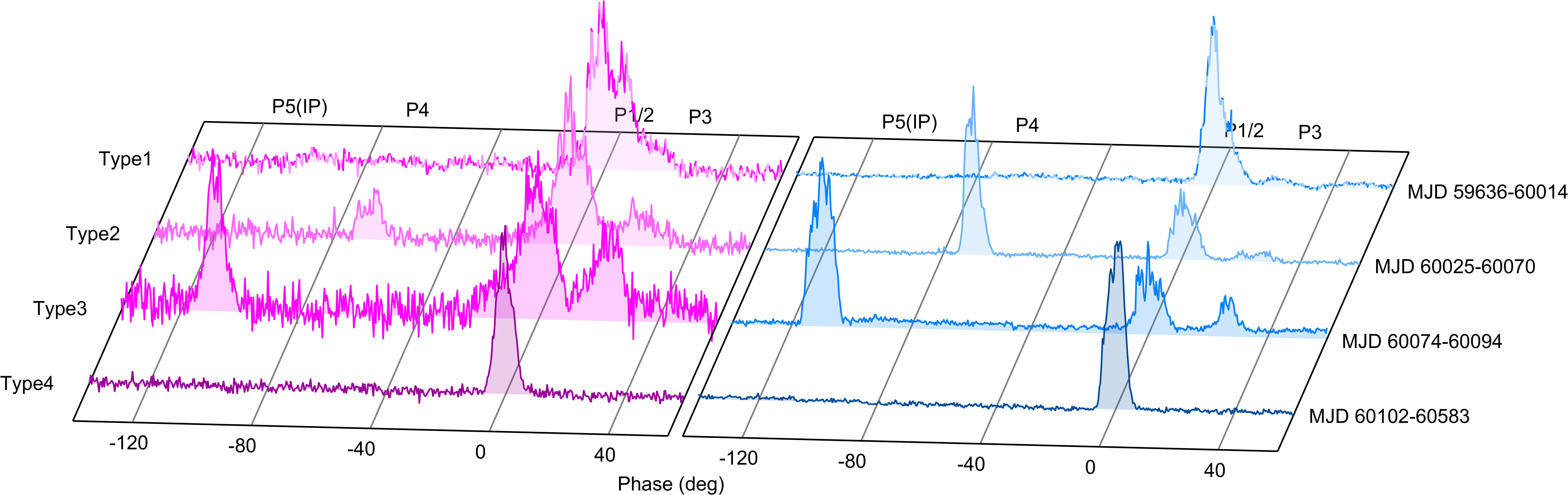}
\caption{Representative integrated profiles of XTE~J1810$-$197 at 2.25/8.60~GHz, showcasing four distinct morphological types. For each type, a representative pair of integrated profiles is displayed in one row: the red profile on the left at 2.25~GHz and the blue profile on the right at 8.60~GHz, both obtained from the same observation. Type labels ($Type~1$–$4$) and durations are indicated to the left and right of the integrated profiles, respectively. Prominent components are labeled \textit{P1}-\textit{P5} and marked above the profiles.
\label{fig:profile_type}}
\end{figure*}

Pulsed radio emission from magnetars is known to be highly variable, with integrated profiles that evolve significantly over time. In our previous work, we confirmed that XTE~J1810$-$197 exhibited twelve distinct types of integrated profiles during 926 days, as illustrated in Figure~2 of Huang et al. (2023) \cite{hys2023}. To investigate the long-term morphological evolution of XTE~J1810$-$197 in the following observations, we analyzed 100 observations with detectable radio emission obtained between MJD~59636 and 60583 (see Figure~S1 in Supplementary Materials for all integrated profile plots). We found that the integrated pulse profiles of XTE~J1810$-$197 underwent significant morphological changes over the course of the outburst. On long timescales, the profiles evolved from broad, multi-component structures (e.g., MJD~59636–60014, where both 2.25~GHz and 8.60~GHz profiles were composed of a central component and a trailing component) to narrower and simpler shapes dominated by a central component (e.g., after MJD~60102). Notably, two leading components sequentially appeared at phases $-72^\circ$ and $-110^\circ$ during MJD~60025–60094 (see Figure~S1 in Supplementary Materials). On shorter timescales, the integrated profiles often exhibited morphological stability, with consistent emission components appearing at similar pulse phases, though their relative amplitudes can vary. For instance, the integrated profile, dominated by a single central component after MJD~60102, remained relatively stable over a long duration of more than 400 days, until the last detection on MJD~60583.

Motivated by this structural stability, we classified the integrated profiles into four distinct morphological types, based on the presence and configuration of active components. To illustrate these morphological types and their temporal evolution, Figure~\ref{fig:profile_type} presents representative examples, with simultaneous observations at 2.25~GHz (red) and 8.60~GHz (blue). From top to bottom, the figure shows four morphological types ($Type~1-4$), each arranged in a separate row. For each type, we selected high S/N integrated profiles from its corresponding time interval, allowing comparison of profile evolution between different types. All profiles are aligned in phase such that the most frequently observed and typically brightest component is located at phase zero, based on phase alignment from timing solutions. To facilitate shape comparisons across epochs and frequencies, all profiles are normalized to their peak.
\begin{itemize}[itemsep=2pt, parsep=0pt, topsep=4pt]
  \item $Type~1$, observed from MJD~59636 to MJD~60014, consists of a prominent central component followed by a distinct trailing component. The central component often shows internal substructure, with two resolvable peaks.
  \item $Type~2$, observed approximately between MJD~60025 and MJD~60070, features the emergence of a distinct leading component located around phase $-72^\circ$. This new component is especially prominent at 8.60~GHz and can occasionally surpass the central component in intensity. The trailing component also becomes more prominent than in $Type~1$.
  \item $Type~3$, detected between MJD~60074 and MJD~60094, is characterized by the appearance of a new emission feature near phase $-110^\circ$. This component is particularly strong at 8.60~GHz, with a typical width at 10\% of the peak intensity ($W_{10}$) $\approx 11.4^\circ$, closely matching the width of the IP (about $12^\circ$) detected during the 2003 outburst using Effelsberg at 8.35~GHz \cite{ksj2007,ljk2008}. The strong agreement in both phase location and width supports its identification as an IP, marking the first definitive detection of IP emission during the 2018 outburst. A possible earlier instance was reported at MJD~58552, where a weak feature was seen at 8.60~GHz without a counterpart at 2.25~GHz, but this detection was marginal. In contrast, the IP in $Type~3$ is significantly stronger and persists in multiple epochs. The IP remains more prominent at 8.60~GHz, suggesting a relatively flat spectral index compared to the main pulse.
  \item $Type~4$, observed from MJD~60102 to the final detection at MJD~60583, exhibits the simplest morphology in the sequence. These profiles consist solely of a central component, with both leading and trailing components no longer detectable. Compared to earlier types, the central component is narrower and still exhibits two closely spaced but resolvable peaks.
\end{itemize}

In summary, four distinct types of integrated pulse profiles were detected over a 947-day interval (MJD~59636-60583). However, during a comparable timespan from MJD~58501 to 59427 (spanning 926 days), twelve different profile types were observed, indicating that the profile morphology of this magnetar became more stable over time \cite{hys2023}.

\begin{figure*}[ht!]
\includegraphics[width=1\textwidth]{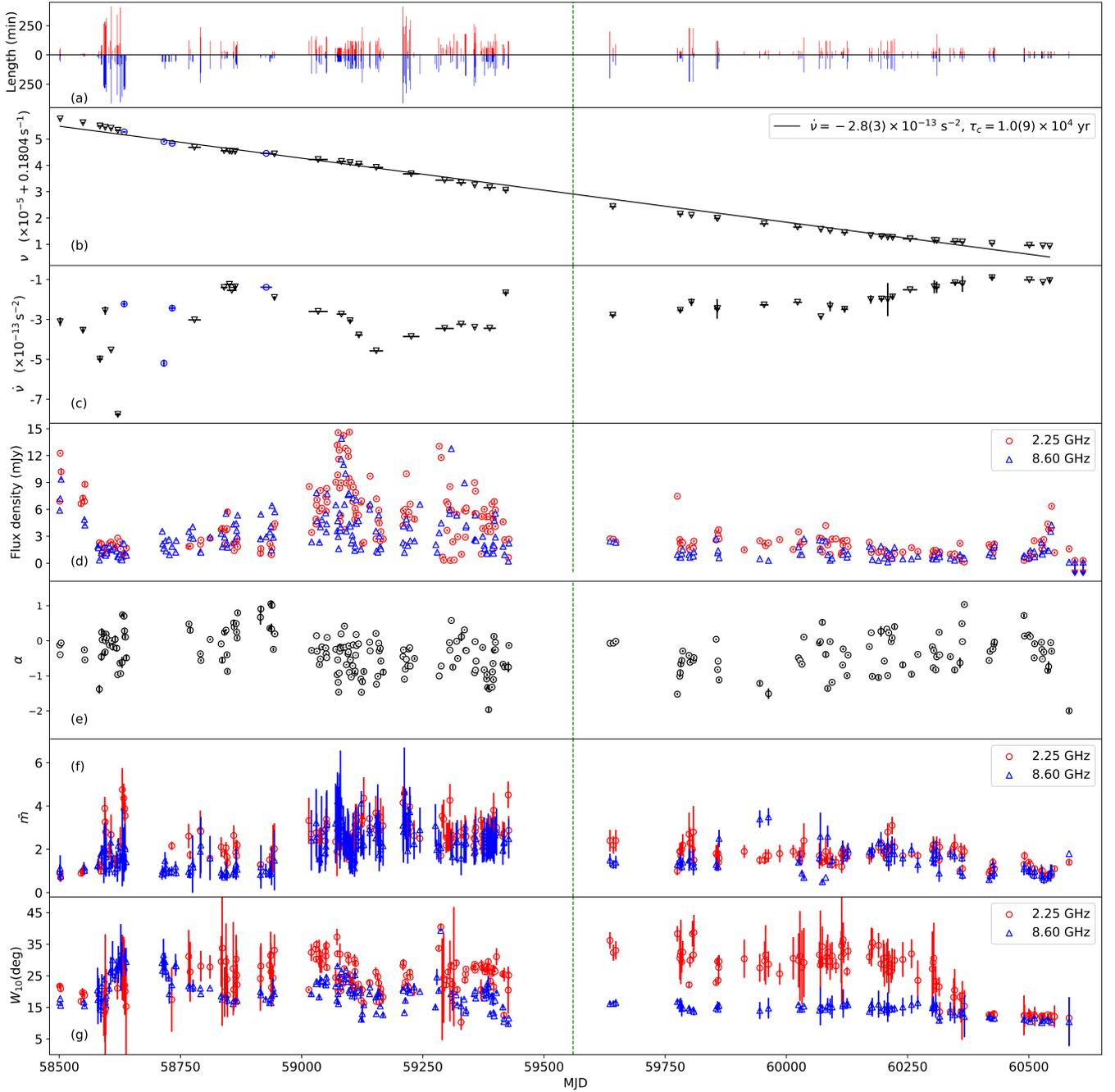}
\caption{
Temporal evolution of observational and derived parameters of XTE~J1810$-$197. 
(a) Red and blue vertical bars show $T_{obs}$ at 2.25/8.60~GHz, respectively, for epochs used in timing analysis. 
(b) $\nu$ evolution, with dual-frequency timing results shown as black downward triangles and 8.60~GHz-only results as blue open circles. The black solid line indicates a linear fit across the active phase. 
(c) $\dot{\nu}$ evolution. 
(d) Flux density measurements at both frequencies, with red open circles and blue open triangles indicating 2.25/8.60~GHz detections, respectively. For the final two observations, flux upper limits are shown as downward arrows. 
(e) Spectral index ($\alpha$) derived from dual-frequency power-law fits. 
(f) Evolution of the phase-resolved modulation index ($\bar{m}$) for the central component ($P1/2$).
(g) Temporal evolution of $W_{10}$ for the $P1/2$ region, determined through Gaussian fitting. 
The vertical green dashed lines separate results from our previous work and those from this study \cite{hys2023}. 
}
\label{fig:para}
\end{figure*}

\subsection{Timing} \label{sec:tim}
Due to the highly variable pulse profiles and unstable spin behavior of XTE~J1810$-$197 \cite{crh2016}, a single phase-connected timing solution could not be established. Instead, we employed a segment-based method similar to previous work \cite{ccr2007,crd2022}, dividing the dataset into shorter segments (spanning 3-15 days) during which the profile morphology and rotational parameters remained relatively stable. In each segment, a standard template was created using the \texttt{paas} tool in \texttt{PSRCHIVE} by selecting the integrated profile with high S/N. The minimum S/N among all the selected integrated profiles across all segments was 14.6. We then derived time-of-arrival (ToA) measurements by cross-correlating individual profiles with the template using the \texttt{pat} command. An example of the typical ToA uncertainties is presented in the sample timing segment in Table~S1, and the number of ToAs for each segment is detailed in Table~S2 of the Supplementary Materials. Dual-frequency ToAs were combined after correcting for instrumental offsets. However, to avoid degradation of timing precision due to low-S/N profiles \cite{tay1992,ehm2006}, we used only the data at 8.60~GHz in segments where the profile at 2.25~GHz had low S/N.

The integrated time of all observations used for timing analysis is shown in Figure~\ref{fig:para}(a), with red and blue vertical bars representing integration times at 2.25/8.60~GHz, respectively. Results derived from 8.60~GHz-only segments are indicated by open blue circles in Figures~\ref{fig:para}(b) and \ref{fig:para}(c). They are also presented in Table~S2 of the Supplementary Materials. The timing parameters for each segment were fitted using \texttt{TEMPO2} \cite{hem2006}, with the segment duration extended iteratively until either significant changes in pulse shape were observed or the uncertainty of $\dot{\nu}$ exceeded 5\%.

The timing results are presented in Figures~\ref{fig:para}(b) and \ref{fig:para}(c), which show the temporal evolution of $\nu$ and its first derivative $\dot{\nu}$, respectively. These plots incorporate both the results from this work and our previously published data, covering the full duration of the 2018 outburst from MJD~58501 to 60583. The two datasets are separated by vertical green dashed lines. The horizontal error bars indicate the ToA span for each timing segment. As shown in Figure~\ref{fig:para}(b), the $\nu$ decreases steadily throughout the outburst, despite the frequent fluctuations observed in $\dot{\nu}$. A linear least-squares fit to all $\nu$ measurements yields a value of approximately $-2.8(3) \times 10^{-13}~\mathrm{s}^{-2}$, corresponding to a characteristic age of $\tau_c \approx 1.0(9) \times 10^4~\mathrm{yr}$.

The evolution of $\dot{\nu}$, as shown in Figure~\ref{fig:para}(c), can be broadly divided into four temporal regimes. Following the initial outburst, from MJD~58502 to 58584, the $\dot{\nu}$ continued to decrease, extending the declining trend previously reported by Levin et al. (2019) during the earliest phase of activity (MJD~58470--58507) at 1.52~GHz \cite{lls2018}. Between MJD~58595 and 58927, $\dot{\nu}$ exhibited strong short-term fluctuations, varying by a factor of 3 at timescales of roughly 10 days. Subsequently, from MJD~58840 to 59154, a more coherent monotonic decrease was observed, with $\dot{\nu}$ dropping from $-1.25(5) \times 10^{-13}~\mathrm{s}^{-2}$ to $-4.58(2) \times 10^{-13}~\mathrm{s}^{-2}$. After MJD~59154, the $\dot{\nu}$ began to recover, gradually returning toward the pre-outburst level of approximately $-1.2 \times 10^{-13}~\mathrm{s}^{-2}$ \cite{pme2019}, and remained relatively stable thereafter. This evolution is broadly consistent with previous observations, and its comparison with the 2003 outburst behavior will be discussed in sect.~\ref{sec:discussion}. 

Rajwade et al. (2022) reported that distinct phases of $\dot{\nu}$ evolution in Swift~J1818.0$-$1607 correlate with different integrated pulse profile shapes \cite{rsl2022}. This spin-down/profile correlation has also been observed in the high-$B$ pulsar PSR~J1119$-$6127 during its magnetar-like active phase \cite{djw2018}. However, our current observations of XTE~J1810$-$197 during its second active phase show a different behavior. Although we detected four distinct integrated pulse profile segments, the source's $\dot{\nu}$ throughout the entire MJD~59636-60612 period remained within a relatively steady recovery trajectory.

According to previous research, the $\dot{\nu}$ of XTE~J1810$-$197 in radio-active phase is known to be highly variable, fluctuating over a wide range from $-7 \times 10^{-13}$ to $-9 \times 10^{-14}~\mathrm{s}^{-2}$ \cite{ims2004,crh2016,lld2019,crd2022}. Caleb et al. (2022) demonstrated this variability by fitting $\dot{\nu}$ in 30-day time windows, which resulted in highly scattered characteristic ages ranging from $400$ to $2.4 \times 10^4$~yr, highlighting the care needed when interpreting $\tau_c$ from short-term data. Given the clear linear decline trend in $\nu$ (see Figure~\ref{fig:para}), we previously adopted a long-term least-squares fit across our initial 194 epochs dataset (MJD~58501–59427), yielding $\dot{\nu}\sim -3.2(1) \times 10^{-13}~\mathrm{s}^{-2}$ and $\tau_c \sim 8.9(2) \times 10^3$~yr. Incorporating the subsequent 100 observations (MJD~58501–60583), our long-term average $\dot{\nu}$ is now $-2.8(3) \times 10^{-13}~\mathrm{s}^{-2}$, corresponding to $\tau_c \sim 1.0(9) \times 10^4$~yr. Crucially, this active-phase value is in stark contrast to the quiescent state, where $\dot{\nu}$ remained stable at approximately $-9.2 \times 10^{-14}~\mathrm{s}^{-2}$ \cite{crh2016,pbm2016,pme2019}, which suggests a much older characteristic age of $\sim 3.1 \times 10^4$~yr.

\subsection{Flux Densities and Spectral Indices}
Based on the method described in sect.~\ref{sec:obs}, we calculated the mean flux densities at 2.25/8.60~GHz (denoted as $S_{2.25}$ and $S_{8.60}$, respectively) for each observation epoch. For the final two observations at MJD~60595 and 60612, where no radio signal was detected, we estimated $3\sigma$ upper limits on flux densities by assuming the same pulse profile shape as that of the last high-S/N detection on MJD~60547, and by adopting a peak S/N of 3. The long-term flux evolution is presented in Figure~\ref{fig:para}(d), which combines the measurements from this work with those from our previous work. In Figure~\ref{fig:para}(d), the flux densities at 2.25/8.60~GHz are shown as red open circles and blue open triangles, respectively. The two upper-limit values are denoted by downward arrows at their respective epochs.

As shown in Figure~\ref{fig:para}(d), both $S_{2.25}$ and $S_{8.60}$ follow similar trends, which can be divided into five stages. Before MJD~58579, both frequencies showed a stable declining trend, consistent with 1.52~GHz measurements from the Lovell telescope \cite{lls2018}. The subsequent $\sim 400$ days featured a sustained low-flux state. Between MJD~59095 and 59427, the source entered a highly variable phase, with flux densities changing by factors of five within several days. This was followed by a prolonged low-emission period lasting $\sim 900$ days (MJD~59636--60527), during which all flux measurements, except the 2.25~GHz point at MJD~59775, remained below 3.53~mJy. In the final three observations in which radio emission was detected (MJD~60541, 60547, 60583), the flux showed brief surges followed by rapid declines. $S_{2.25}$ remained comparable to previous levels, while the $S_{8.60}$ dropped significantly. Radio emission at both frequencies became undetectable just 12 days later on MJD~60595. We conducted 30~min observations on both MJD~60595 and MJD~60612, but no signal was detected at either 2.25~GHz or 8.60~GHz. The corresponding flux density upper limits were 0.3~mJy and 0.1~mJy, respectively.

Spectral indices derived from simultaneous dual-frequency observations are plotted in Figure~\ref{fig:para}(e). While considerable variability is present, the source generally maintained a flat radio spectrum. In approximately 90\% of the epochs (206 of 233), the spectral index was approximately flat, with $\alpha > -1$. This flat-spectrum behavior persisted into the final stage of radio activity, with a notable exception at MJD~60583, when the index steepened to $\alpha = -2.01$. In addition, we measured the spectral indices of the IP during three epochs---MJD~60074, 60090, and 60094---obtaining values of $0.54(6)$, $-0.61(3)$, and $-0.36(3)$, respectively. Although the spectral indices of the IP vary significantly between these epochs, its spectrum consistently remained flatter than that of the corresponding MP, consistent with observations obtained during the 2003 outburst \cite{ksj2007,ljk2008}.

\section{Comparison and Discussion} \label{sec:discussion}
\subsection{Pulse Profile Evolution and Spiky Emission}

XTE~J1810$-$197 continued to exhibit evolving integrated pulse profile shapes throughout the current monitoring campaign. During the 902-day interval from MJD~59636 to 60583, we conducted 100 dual-frequency (2.25/8.60~GHz) observations and identified four distinct morphological types. By contrast, our previous work covering a comparable 926-day span revealed twelve distinct profile types \cite{hys2023}. This comparison suggests that although the pulse profile remained variable, the degree of morphological diversity had diminished significantly during the later stages of the outburst. A similar trend was reported during the 2003 outburst: Camilo et al. (2016) noted that the profiles became less variable in its final phase (2007--2008) \cite{crh2016}.

Similar to other radio-loud magnetars, XTE~J1810$-$197 often exhibits integrated profiles that appear ``rough'' even at high S/N \cite{lbb2012,crh2016}. This apparent roughness is primarily attributed to the presence of narrow, transient bursts of emission---commonly referred to as spiky pulses---which introduce sharp, short-timescale structures into the integrated profile. To quantify the degree of such fine-scale structure, we employed the phase-resolved modulation index $m_i$, which characterizes intensity fluctuations as a function of pulse phase on short timescales \cite{bac1970,wes2006}. For each phase bin $i$, the modulation index is defined as $m_i = \sigma_i / \mu_i$, where $\mu_i$ and $\sigma_i$ are the mean and standard deviation of the flux across sub-integrations (30~s in our observations). Given that the central component remains consistently present across all epochs, we computed the average modulation index $\bar{m}$ within this region to quantify the temporal evolution of the spiky emission state. As shown in Figure~\ref{fig:para}(f), two periods exhibit distinctly elevated average phase modulation index: from MJD~58594 to 58636 at 2.25~GHz, and from MJD~59015 to 59427 at both 2.25 and 8.60~GHz. The rise observed at 2.25~GHz between MJD~58594 and 58636 is likely influenced by the relatively low signal-to-noise ratio of the profiles during that interval. Since phase-resolved modulation indices tend to be overestimated in weak emission regions, this effect may inflate the values. Notably, no corresponding increase was seen at 8.60~GHz, where the emission remained stronger. These results suggest that although the modulation index remained elevated compared to ordinary pulsars, the profiles of XTE~J1810$-$197 were relatively smoother during the early and late stages of the outburst.

To further quantify the structural evolution of the integrated pulse profiles during this phase of the outburst, we applied Gaussian fitting to all profiles obtained between MJD~59636 and 60583. The Gaussian fitting was performed using a standard least-squares method. Based on the phase location and width of Gaussian components, we identified five distinct components, labeled $P1$ through $P5$, as marked in Figure~\ref{fig:profile_type}. Our Gaussian analysis confirms that the central emission structure consistently separated into two adjacent components, labeled $P1$ and $P2$, in agreement with visual inspection. Given the persistent presence of this combined $P1$/$P2$ region in all epochs, we calculated its $W_{10}$ based on the fitted Gaussian models. These measurements, combined with historical data, are plotted in Figure~\ref{fig:para}(g). As shown in the figure, the $P1$/$P2$ region at 2.25~GHz is generally wider than at 8.60~GHz, except during the interval from MJD~58615 to 58741, where the widths are comparable or reversed. After MJD~60271, both frequency bands exhibit a clear narrowing trend in $P1$/$P2$ width, converging to approximately $11^\circ$ after MJD~60362.

\subsection{Comparison with the 2003 and 2018 Outbursts}
\begin{figure*}[ht!]
\includegraphics[width=0.95\textwidth]{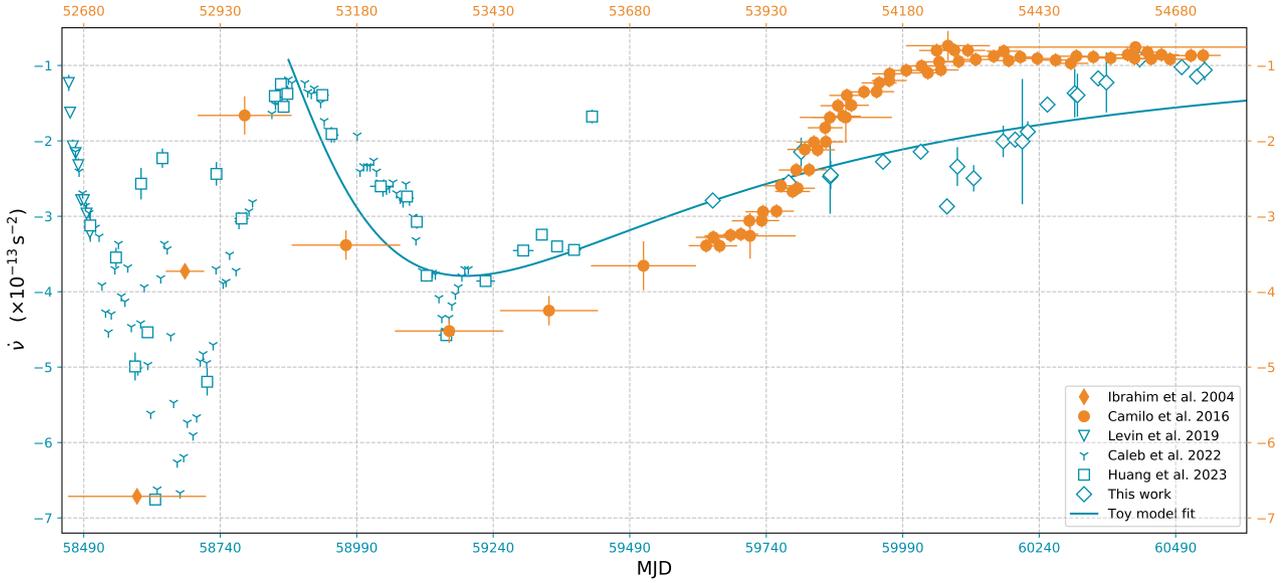}
\caption{Comparison of the $\dot{\nu}$ evolution of XTE~J1810$-$197 during its 2003 and 2018 outbursts. 
The $\dot{\nu}$ values from the 2003 outburst (orange) and 2018 outburst (light blue) are aligned such that the epochs of radio disappearance coincide. Orange diamonds are from Ibrahim et al. (2004) \cite{ims2004}, orange filled circles are from Camilo et al. (2016) \cite{crh2016}, light blue open triangles are from Levin et al. (2019) \cite{lld2019}, light blue triangles are from Caleb et al. (2022) \cite{crd2022}, light blue open squares are our previous work \cite{hys2023}, and light blue open diamonds represent results from this work.
\label{fig:F1compare}}
\end{figure*}

In order to compare the evolution of the $\dot{\nu}$ at the two distinct active phases of XTE~J1810$-$197, we present the $\dot{\nu}$ measurements from both the 2003 and 2018 outbursts in Figure~\ref{fig:F1compare}, shown in orange and light blue, respectively. Due to the sparse observational coverage during the first $\sim$100 days following the 2003 X-ray outburst, we aligned the two datasets by matching the epochs of radio disappearance, thereby facilitating a clearer comparison of their post-outburst evolution. Despite the limited early-time data for the 2003 outburst, the two available measurements fall within the same magnitude range as those observed during the corresponding phase of the 2018 event \cite{ims2004}. More notably, after $\sim$500 days post-outburst in both cases, $\dot{\nu}$ displays a strikingly similar evolutionary pattern. In each outburst, $\dot{\nu}$ initially undergoes a steady decrease over a span of $\sim$400 days, reaching a minimum near $-4.4 \times 10^{-13}$~s$^{-2}$ \cite{crd2022,hys2023}. This is followed by a gradual recovery phase during which $\dot{\nu}$ increases toward a value of approximately $-1 \times 10^{-13}$~s$^{-2}$---comparable to the quiescent level inferred from X-ray timing \cite{pbm2016,pbm2016b}. Finally, $\dot{\nu}$ remains stable at this level until the radio emission ceases \cite{crh2016}. This close agreement in both amplitude and timescale between the two radio-active episodes suggests that a similar physical mechanism governs the long-term $\dot{\nu}$ evolution during each outburst.

In our monitoring campaign, we found that the radio emission from XTE~J1810$-$197 during the 2018 outburst ended abruptly rather than gradually. A similar behavior was reported during the end of the 2003 outburst. As noted by Camilo et al. (2016) \cite{crh2016}, the magnetar ceased its radio emission without any prior indication from its flux or timing properties. In fact, high-quality profiles were still recorded in the weeks leading up to the final detection on 2008 November 3, when the source was observed at 1.4~GHz using Parkes telescope. However, just one week later, on 2008 November 10, no radio emission was detected, and the source remained undetectable in all subsequent radio observations despite extensive follow-up over the next several years. In our final detection at MJD~60583, the flux at 2.25~GHz was measured at $S_{2.25} = 1.60(2)$~mJy, a value comparable to the previous result. In sharp contrast, the simultaneous flux at 8.60~GHz had dramatically declined to $S_{8.60} = 0.11(5)$~mJy, representing a decrease by a factor of $\sim 10$ compared to its previous value. Our simultaneous dual-frequency coverage at both 2.25 and 8.60~GHz provides further evidence that the high-frequency emission may have faded earlier than the low-frequency emission.

In addition to similarities in timing and flux evolution, the large-scale magnetospheric geometry of XTE~J1810$-$197 also appears to have remained stable across both outbursts \cite{dwg2024}. Reanalysis of polarization data from the 2006 Effelsberg observations yielded a viewing angle of $\zeta \approx 155^\circ$, in agreement with the value of $\zeta \approx 168^\circ$ derived during the 2018 outburst. Moreover, the shape of the position angle (PA) swing showed no significant variation between the two outbursts. These results indicate that, despite the decade-long interval, the magnetosphere retained a relatively stable global configuration throughout both active phases.

By comparing the evolution of parameters such as integrated profile, $\dot{\nu}$, flux density, spectral index and polarization across the two outbursts of XTE~J1810$-$197, we identify two salient features. First, each radio-active episode concludes with a set of distinctive changes: the integrated profile narrows, the $\dot{\nu}$ recovers to quiescent levels, the radio emission disappears abruptly and the spectrum steepens. Second, the two outbursts exhibit remarkable similarities in multiple parameters, most notably in the temporal evolution of $\dot{\nu}$. The recurrence of such consistent trends, despite the decade-long interval between the events, suggests a repeatable physical process that governs the magnetospheric evolution of XTE~J1810$-$197.

\subsection{Comparative Evolution of Radio-Loud Magnetars and High-$B$ Pulsars}
\begin{table*}[ht!]
	\scriptsize
	\caption{Comparative analysis of evolutionary characteristics among radio-loud magnetars and high-$B$ pulsar.}
	\begin{tabular*}{\textwidth}{lccccccc}
		\toprule
		\multicolumn{1}{c}{Source} & \multicolumn{2}{c}{Radio-active history} & \multicolumn{3}{c}{$\dot{\nu}$ evolution in radio-active phase} & \multicolumn{2}{c}{Late-phase behavior} \\
		\cmidrule(lr){2-3} \cmidrule(lr){4-6} \cmidrule(lr){7-8}
		\multicolumn{1}{c}{Name} & Cessation? & Recurrent? &  Decreased Initially? & Recovery? &  Anticorrelated with flux? & Profile narrowing? & Spectrum steepening? \\
		\midrule
		XTE~J1810$-$197 & Y & Y & Y & Y & Y & Y & Y \\
		1E~1547.0$-$5408 & Y & Y & Y & U & U & U & U \\
		PSR~J1622$-$4950 & Y & Y & Y & Y & Y & Y & U \\
		SGR~J1745$-$2900 & N & N & Y & N/A & U & N/A & N/A \\
		SGR~J1935+2154 & Transient & N/A & N/A & N/A & N/A & N/A & N/A \\
		Swift~J1818.0$-$1607 & N & N & N & N/A & Y & N/A & N/A \\
		\textbf{PSR J1119$-$6127} & \textbf{Y} & \textbf{N} & \textbf{Y} & \textbf{Y} & \textbf{Y} & \textbf{Y} & \textbf{U} \\
		\bottomrule
	\end{tabular*}
	\begin{tablenotes}
		\item[1)]\textbf{Notes.} Radio-loud high-$B$ pulsar is labeled with the bold font. The comparative results are divided into Yes (Y), No (N), Unknown or insufficient data (U) and Not Applicable (N/A). ref:\cite{crh2007, mgw2009, lbb2010, sk2011, dks2012, efk2013, ier2016, crh2016, scs2017,css2018, lld2019, rcv2020, hys2021, rsl2022, lys2023, hne2024}
			\label{tab:evolution_compare}
	\end{tablenotes}
\end{table*}

Our analysis of XTE~J1810$-$197's two radio-active phases reveals significant similarities in its spin-down, profile, and spectral evolution, suggesting a consistent, underlying physical process governing its magnetospheric activity. To determine whether this evolutionary pattern is unique to XTE~J1810$-$197 or indicative of a universal behavior, we extend our discussion to include a comparative analysis of other radio-emitting magnetars and the high magnetic field (high-$B$) pulsar PSR~J1119$-$6127, which possesses pulsed radio emission and exhibits magnetar-like behavior. The comparison, summarized in Table~\ref{tab:evolution_compare}, focuses on key evolutionary metrics during the radio-active phase. The sources listed in the table are categorized into two groups: radio-loud magnetars (Rows 1--6: XTE~J1810$-$197, 1E~1547.0$-$5408, PSR~J1622$-$4950, SGR~J1745$-$2900, SGR~J1935+2154, and Swift~J1818.0$-$1607) and the high-$B$ pulsar PSR~J1119$-$6127 (Row 7, distinguished by bold font). The magnetars are listed in approximate chronological order based on the order in which their pulsed radio emission was discovered. The comparative results are presented qualitatively, with 'Y' denoting "Yes," 'N' denoting "No," 'U' indicating "Unknown or insufficient data," and 'N/A' for "Not Applicable." SGR~J1935+2154 is unique within this sample, as it only exhibits transient radio bursts and lacks the sustained pulsed emission required for long-term timing and spectral analysis \cite{ier2016,hne2024}. Consequently, the evolutionary metrics are denoted as 'N/A' in Table~\ref{tab:evolution_compare}.

Generally, the pulsed radio emission from magnetars is observed following an X-ray outburst and typically remains active over a timescale of several years. XTE~J1810$-$197, for instance, showed the radio-active phase lasting approximately six years \cite{crh2016}. The radio histories of the classical transient magnetars in our sample exhibit recurrent and intermittent behavior: 1E~1547.0$-$5408, first detected in radio in 2007 \cite{crh2007}, showed intermittent radio pulsations following subsequent dramatic X-ray bursts in 2008 and 2009 \cite{mgw2009,sk2011}. While it maintained continuous radio emission from approximately 2013 onward, it experienced a brief cessation of pulsed radio signal in 2022, which was contemporaneous with an X-ray burst \cite{lys2023}. PSR~J1622$-$4950 has reportedly experienced at least three distinct radio-active phases: 2000--2002, 2009--2014, and the latest one beginning around 2017 \cite{lbb2010,scs2017,css2018}. Unfortunately, precise onset and termination times for the earlier two phases are difficult to ascertain due to limited data coverage. No recent published reports indicate that the source has ceased its radio emission since the 2017 re-activation. In contrast, the more recently discovered radio-loud magnetars have not yet been reported to cease their emission: SGR~J1745$-$2900, discovered via an X-ray burst in 2013 \cite{efk2013}, was reported to remain radio-active as late as 2020 \cite{dem2025}. Similarly, Swift~J1818.0$-$1607, discovered in 2020 \cite{egk2020}, has not yet shown signs of complete radio turn-off in published literature. While magnetars can sustain their radio activity for years, the new magnetar-like radio emission state of PSR~J1119$-$6127, which appeared following its 2016 burst, lasted for a much shorter duration, approximately 100 days \cite{djw2018}.

The $\dot{\nu}$, which reflects the external magnetic braking torque, provides the most consistent signature of magnetar-type activity across the sample. For most sources with sufficient timing data (including XTE~J1810$-$197, 1E~1547.0$-$5408, SGR~J1745$-$2900, and the high-$B$ pulsar PSR~J1119$-$6127), a significant initial decrease in $\dot{\nu}$ was measured \cite{dks2012,djw2018,lld2019,rcv2020}. However, Swift~J1818.0$-$1607 stands as a notable exception to this initial pattern; its $\dot{\nu}$ does not show a clear initial drop, but rather exhibits highly variable, short-term oscillations in the $\dot{\nu}$ from the beginning of the radio-active phase \cite{hys2021,rsl2022}. Following the initial phase, XTE~J1810$-$197 showed a clear trend of $\dot{\nu}$ recovery. This recovery phase is a shared trait, also definitively observed in PSR~J1622$-$4950 and PSR~J1119$-$6127 \cite{scs2017,djw2018}, where $\dot{\nu}$ moved toward the pre-outburst quiescent level before the radio cessation. Furthermore, the link between the magnetospheric structure and the radio emission mechanism is established by the anticorrelation between $\dot{\nu}$ and radio flux in XTE~J1810$-$197 \cite{crh2016}. This exact anticorrelation is explicitly noted in both PSR~J1622$-$4950 and PSR~J1119$-$6127 \cite{scs2017,djw2018}. Additionally, for Swift~J1818.0$-$1607, this physical connection is established through its oscillating $\dot{\nu}$ being anticorrelated with distinct pulse profile modes \cite{rsl2022}.

The radio emission properties in the phase immediately preceding the radio cessation event offer critical constraints on the mechanism that terminates the radio activity. In the late stages of XTE~J1810$-$197's radio activity, this magnetar showed both pulse profile narrowing and evidence for abrupt spectral steepening just prior to the final radio disappearance \cite{crh2016}. However, due to the limited archival observational data near cessation events, a comparison with other sources is constrained. Nevertheless, the pulse profile narrowing observed in XTE~J1810$-$197 is also definitively documented as a late-phase feature for PSR~J1622$-$4950 and the high-$B$ pulsar PSR~J1119$-$6127 \cite{scs2017,djw2018}. Data on late-stage spectral steepening are currently insufficient for other sources, preventing a conclusion on its universality.

\subsection{Magnetospheric Untwisting Model and Comparative Insights}

As discussed above, the $\dot{\nu}$ evolution of radio-loud magnetars exhibits remarkable similarities, suggesting a universal underlying physical mechanism. To explore this mechanism, we introduce the simplified magnetospheric twist decay model (toy model) \cite{th2020}. The rationale for its use is twofold. Firstly, this model successfully provides a unified physical explanation for the observed enhanced spin-down immediately following an outburst and the subsequent gradual $\dot{\nu}$ recovery \cite{th2020}. Secondly, the model is directly motivated by geometric and timing analysis of this specific source: Desvignes et al. (2024) found that the radio polarization variations of XTE~J1810$-$197 were best interpreted as free precession coupled with a slowly untwisting magnetosphere \cite{dwg2024}. In this work, the toy model supported the free-precession interpretation. Hence, we employed the toy model based on the untwisting magnetosphere framework to explain the torque evolution during magnetar outbursts \cite{th2020}. In this model, the effective magnetic field strength $B(t)$ evolves as a combination of local twist accumulation (which increases $B$) and global twist decay (which decreases $B$), leading to the expression:

\begin{equation}
\frac{B(t)}{B_0} = 1 + A(1 - e^{-t/\tau_1})e^{-t/\tau_2},
\end{equation}

where $B_0$ is the initial untwisted field strength, $A$ is an amplitude factor, and $\tau_1$ and $\tau_2$ are characteristic timescales for the local and global twist evolution, respectively. Assuming the spin-down torque scales with the square of the magnetic field strength, i.e., $\dot{\nu}(t) \propto B(t)^2$, the corresponding evolution of $\dot{\nu}$ is given by:

\begin{equation}
\frac{\dot{\nu}(t)}{\dot{\nu}_0} = \left[1 + A(1 - e^{-t/\tau_1})e^{-t/\tau_2} \right]^2.
\end{equation}

We applied this model to the decay phase of the 2018 outburst of XTE~J1810$-$197, focusing on the interval from MJD~58864 to 60583, where the $\dot{\nu}$ evolves more smoothly after the initial rapid variations. By fitting the model to the observed $\dot{\nu}$ using a least-squares method, we obtained best-fit parameters of $A = 1.74$, $\tau_1 = 178.2$ days, and $\tau_2 = 920.6$ days. The fitted curve is plotted in Figure~\ref{fig:F1compare} with light blue solid line, showing agreement with the data during this stage. A previous study interpreting the polarization variations of the magnetar's radio emission employed a combined framework of free precession and magnetospheric untwisting \cite{dwg2024}. In that work, the inferred relaxation timescale exceeds 730 days, which is consistent with the global untwisting timescale of 926 days derived from our model fit.

The untwisting magnetosphere model also provides a coherent physical explanation for the evolution and eventual disappearance of radio emission in XTE~J1810$-$197. In this model, a twisted bundle of current-carrying magnetic field lines---often referred to as the $j$-bundle---gradually contracts toward the magnetic axis over the course of the outburst \cite{dt1992,td1995,b2009}. This contraction leads to a shrinking open-field region, resulting in narrower integrated pulse profiles and reduced radio flux \cite{b2009}. To address the sudden disappearance of radio emission after the two outbursts and the prior fading of high-frequency radiation, we refine the explanation as follows:
\begin{itemize}
\item \textbf{Explanation for the Sudden Disappearance of Radio Emission:} The magnetosphere untwisting model suggests that the current-carrying region ($j$-bundle) contracts towards the magnetic axis over time. This reduces both the volume of the emitting region and the angular extent of open magnetic field lines, leading to narrower pulse profiles and weaker radio emission. The radio emission ceases when the $j$-bundle contracts to the point where the magnetic field strength at the light cylinder drops below the threshold required for radio waves to escape the magnetosphere.
\item \textbf{Explanation for the Prior Fading of High-Frequency Radio Emission:} In a plasma-filled magnetosphere, the ability of radio waves to escape depends on their emission location relative to a threshold magnetic surface. High-frequency radiation fades before low-frequency radiation because the escape condition is more difficult to satisfy for higher frequencies. The high-frequency radiation is emitted from regions closer to the magnetar's surface, where the magnetic field strength is higher. As the magnetosphere untwists and the magnetic field strength decreases, the high-frequency radiation cannot escape once the magnetic field drops below the critical threshold, leading to the prior fading of high-frequency emission.
\end{itemize}

\begin{figure*}[ht!]
\includegraphics[width=0.95\textwidth]{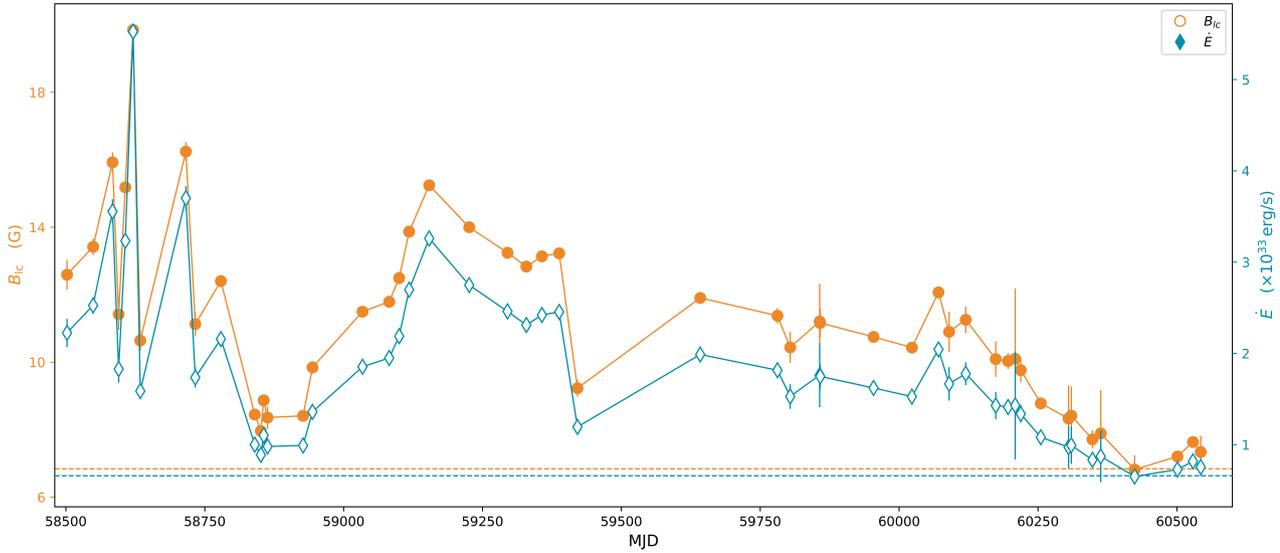}
\caption{Evolution of $B_{\text{lc}}$ (orange solid line with circles) and spin-down luminosity $\dot{E}$ (light blue solid line with open diamonds) of XTE~J1810$-$197. The orange and light blue dashed lines mark the corresponding values of $B_{\text{lc}}$ and $\dot{E}$ during the magnetar's quiescent state, respectively.}
\label{fig:BLc_Bsurf_Edot}
\end{figure*}

This interpretation is further supported by the temporal evolution of several physical parameters: the magnetic field at the light cylinder $B_{\mathrm{lc}} = B_{\mathrm{surf}} (R/R_{\mathrm{lc}})^3$, and the spin-down luminosity $\dot{E} = 4\pi^2 I \dot{P}/P^3$. Here, $P$ is the spin period, $\dot{P}$ its derivative, $I$ the moment of inertia (assumed to be $10^{45}$~g~cm$^2$), and $R$ the neutron star radius (10~km). As shown in Figure~\ref{fig:BLc_Bsurf_Edot}, the evolution of $B_{\mathrm{lc}}$ and $\dot{E}$ shows some complexity over the course of the outburst, all parameters exhibit a converging trend toward their quiescent values in the final stages. This convergence suggests that the magnetosphere had relaxed to an untwisted configuration.

While the untwisting magnetosphere model provides a comprehensive framework for interpreting the observed radio emission behavior, other alternative models could also potentially explain the phenomena observed in this paper. The free precession model suggests that slow, periodic changes in the viewing geometry---due to misalignment between the magnetar’s spin and symmetry axes---could modulate radio visibility \cite{dwg2024}. Alternatively, the two-stream instability model proposes that plasma instabilities in the magnetosphere could rapidly disrupt coherent emission \cite{aro1983}. 

Beyond its application to magnetar outbursts, the magnetospheric untwisting model has also been proposed as a unifying framework for explaining the on--off switching behavior in intermittent pulsars \cite{hyt2016}. In both scenarios, localized or global magnetic twists in the closed field-line region near the neutron star surface drive electric currents that enhance spin-down torque. As the twisted field dissipates, the magnetosphere relaxes to a lower-energy state, leading to reduced radio activity or complete quenching of coherent emission. This theory alignment suggests a potential physical connection between radio magnetars and intermittent pulsars, despite differences in surface magnetic field strength or outburst energetics. In particular, while intermittent pulsars are thought to undergo local magnetospheric reconfiguration, radio-loud magnetars like XTE~J1810$-$197 likely experience more global-scale untwisting events. These differences in spatial scale should naturally lead to different levels of torque variation between active and quiescent phases.

\begin{figure*}[ht!]
\centering
\includegraphics[width=0.65\textwidth]{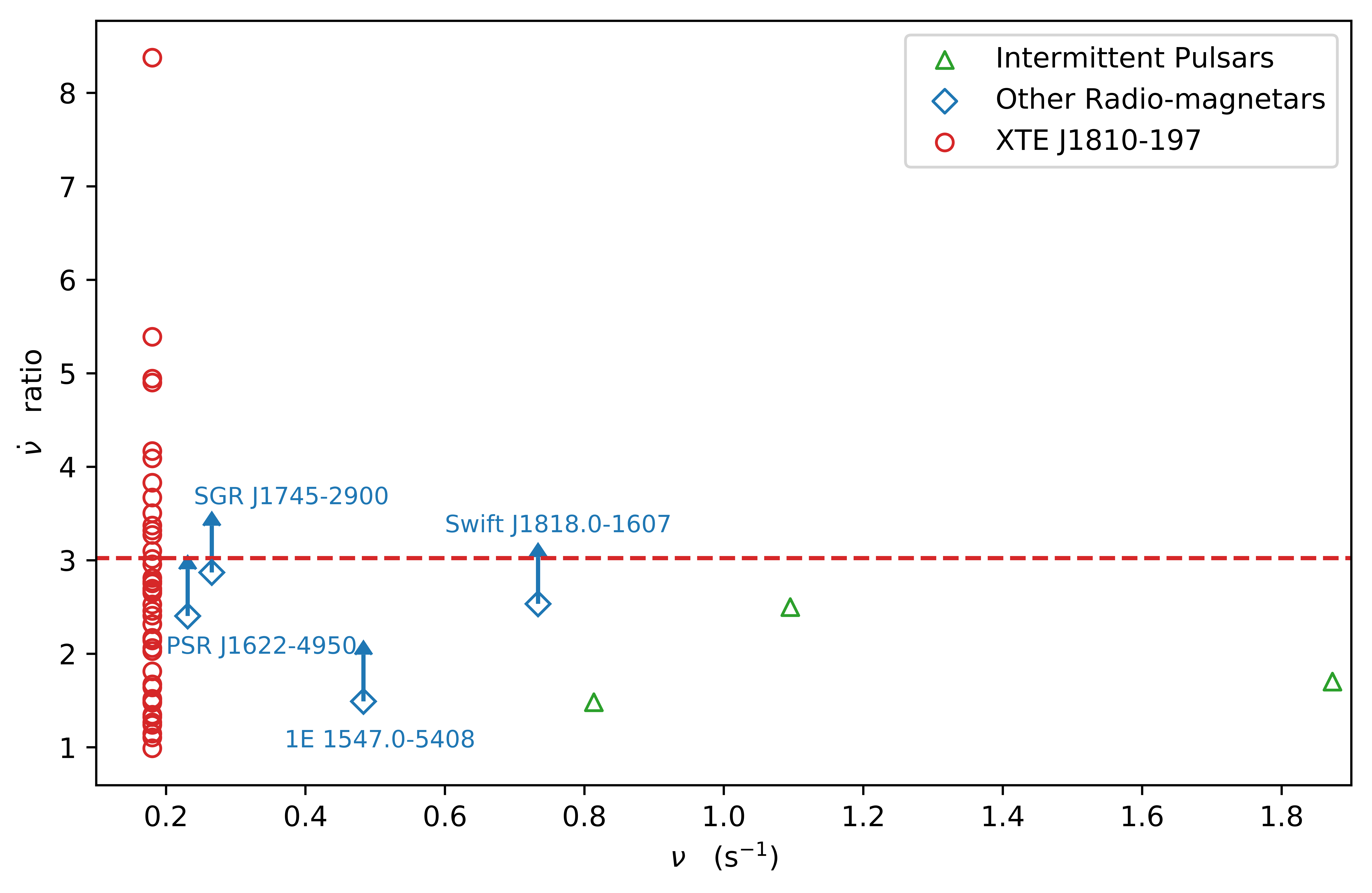}
\caption{Spin-down rate ratio $\dot{\nu}$ (on/off) for intermittent pulsars (green triangles), other radio magnetars (light blue diamonds), and XTE~J1810$-$197 (coral circles). The red dashed line indicates the spin-down rate ratio of XTE~J1810$-$197, derived from the average $\dot{\nu}$ across its entire radio-active period compared to its quiescent state \cite{scs2017,hys2021,rcv2020,lys2023}. }
\label{fig:F1_ratio}
\end{figure*}

According to previous theoretical calculations \cite{hyt2016}, the enhancement in spin-down rate caused by magnetospheric reconfiguration depends on the size of the region containing twisted closed field lines. For localized twists, as expected in intermittent pulsars, the additional torque increases $\dot{\nu}$ by a factor of approximately 2 during the radio-active “on” state compared to the radio-quiet “off” state. This prediction is consistent with observationally inferred $\dot{\nu}$ ratios in known intermittent pulsars. 

In contrast, the global-scale untwisting expected in radio-loud magnetars, involving large-scale current systems and extended twisted regions, should lead to more substantial changes in spin-down torque across active and quiescent phases. This expectation is supported by observations of XTE~J1810$-$197. As shown in Figure~\ref{fig:F1_ratio}, intermittent pulsars (green triangles) typically exhibit $\dot{\nu}$ ratios about 2, consistent with localized twisting near the stellar surface. In comparison, XTE~J1810$-$197 (coral dashed line), whose ratio is derived from the average $\dot{\nu}$ during its radio-active phase relative to the quiescent value, exhibits a significantly higher value of 3.06, indicating a more extensive restructuring of the magnetosphere. For other radio magnetars, due to the lack of quiescent-phase timing solutions, we adopt the highest observed $\dot{\nu}$ during their active phases as proxies for their quiescent values, and plot the lower limits on the $\dot{\nu}$ ratio as light blue diamonds. These lower limits also generally exceed the $\dot{\nu}$ ratios observed in intermittent pulsars.

Quasi-periodic on–off transitions have been observed in some intermittent pulsars, often attributed to long-term magnetospheric instabilities or crustal activity. In this context, the strikingly similar behaviors observed during the 2003 and 2018 outbursts of XTE~J1810$-$197 in timing, pulse morphology, flux, and spectral evolution---raise the intriguing possibility that this magnetar may also undergo quasi-periodic radio outbursts.

\section{Conclusions} \label{sec:conc}
In this work, we have presented a dual-frequency monitoring campaign of the radio magnetar XTE~J1810$-$197 during its 2018 outburst, providing a detailed characterization of its timing, flux, spectral, and pulse profile evolution. By combining new observations with previous data, we were able to construct a continuous view of the spin-down evolution and radiative behavior, enabling a comprehensive comparison with the 2003 outburst.

Our results reveal several noteworthy features. First, we confirm that the degree of pulse profile variability diminishes in the late stages of the outburst, and we quantify the spiky emission state using phase-resolved modulation indices in 2018 outburst. Second, we provide dual-frequency evidence that high-frequency radio emission fades prior to low-frequency emission, which precedes the sudden cessation of the magnetar's radio activity. Third, by applying a toy model of the untwisting magnetosphere, we are able to interpret both the long-term torque evolution and the abrupt disappearance of radio emission, linking the observed behavior to the contraction of the current-carrying $j$-bundle and the relaxation of the global magnetosphere. Finally, a comparative analysis with the 2003 outburst demonstrates remarkable consistency in the temporal evolution of $\dot{\nu}$, flux density, spectral index, and large-scale magnetospheric geometry, suggesting a repeatable, underlying physical mechanism governing radio-active episodes in XTE~J1810$-$197.

Looking forward, our results provide a possibility for predicting the timing and radiative properties of future outbursts, and highlight the potential of multi-frequency monitoring to probe the magnetospheric dynamics of radio-loud magnetars. Extending similar studies to other radio-active magnetars and connecting these observations to intermittent pulsars may offer further insights into the universal role of magnetospheric untwisting.

\Acknowledgements{This work was supported in part by the National SKA Program of China (Grant No.~2020SKA0120104), the National Key R$\&$D Program of China (Grant No. 2022YFA1603104), the National Natural Science Foundation of China (Grant Nos. U2031119, 12041301, 1257030658, 12573052, 12573103) and Scientific Research Foundation of Hubei University of Education for Talent Introduction (No. ESRC20240007). We also thank Professor Xiaoyu Lai for insightful comments and all the staff of the TMRT for their support.}
\InterestConflict{The authors declare that they have no conflict of interest.}

\newcommand{\apj}{Astrophys. J.}
\newcommand{\apjl}{Astrophys. J. Lett.}
\newcommand{\apjs}{Astrophys. J. Suppl. S}
\newcommand{\mnras} {Mon. Not. R. Astron. Soc.}
\newcommand{\pasa} {Publ. Astron. Soc. Aust.}
\newcommand{\aap} {Astronom. Astrophys.}
\newcommand{\nat} {Nature}
\newcommand{\aj} {Astron. J.} 
\newcommand{\araa} {Annu. Rev. Astron. Astrophys.}
\newcommand{\ptr}{Phil. Trans. R. Soc. Lond. Ser. A}
\newcommand{\chjas}{Chin. J. Astronom. Astrophys. Suppl.}
\newcommand{\art}{Astronom. Res. Technol.}
\newcommand{\pasp}{Publ. Astron. Soc. Pac.}
\newcommand{\jgrsp}{J. Geophys. Res.: Space Phys.}
\newcommand{\ajp}{Aust. J. Phys.}


\end{multicols}
\end{document}